\documentclass[conference]{IEEEtran}
\IEEEoverridecommandlockouts

\usepackage{cite}
\usepackage{amsmath,amssymb,amsfonts}
\usepackage{algorithmic}
\usepackage{graphicx}
\usepackage{textcomp}
\usepackage{xcolor}
\def\BibTeX{{\rm B\kern-.05em{\sc i\kern-.025em b}\kern-.08em
    T\kern-.1667em\lower.7ex\hbox{E}\kern-.125emX}}

\usepackage{todonotes}
\usepackage{comment}

\begin{document}

%
\title{Chimbuko: A Workflow-Level Scalable Performance Trace Analysis Tool 
\thanks{Exascale Computing Project (17-SC-20-SC)}}

%
%

\author{\IEEEauthorblockN{Sungsoo Ha, Wonyong Jeong, Gyorgy Matyasfalvi,}
\IEEEauthorblockA{
\textit{Brookhaven National Lab.}\\
Upton, USA \\
\{sungsooha,wyjeong,gmatyasfalvi\}@bnl.gov}
\and
\IEEEauthorblockN{Cong Xie}
\IEEEauthorblockA{
\textit{Stony Brook University}\\
Stony Brook, USA \\
cong.xie@stonybrook.edu}
\and
\IEEEauthorblockN{ Kevin Huck}
\IEEEauthorblockA{
\textit{University of Oregon}\\
Eugene, USA \\
khuck@cs.uoregon.edu}
\and
\IEEEauthorblockN{Jong Youl Choi}
\IEEEauthorblockA{
\textit{Oak Ridge National Lab.}\\
Oak Ridge, USA \\
choij@ornl.gov}
\and
\IEEEauthorblockN{
Abid Malik, Li Tang,  Hubertus Van Dam, Line Pouchard, \\ Wei Xu, Shinjae Yoo, Nicholas D'Imperio, Kerstin Kleese Van Dam}
\IEEEauthorblockA{
\textit{Brookhaven National Lab.}\\
\{amalik,ltang,hvandam,pouchard,xuw,sjyoo,dimperio,kleese\}@bnl.gov}
}

%
\maketitle

%
\begin{abstract} 
Because of the limits input/output systems currently impose on high-performance computing systems, a new generation of workflows that include online data reduction and analysis is emerging. Diagnosing their performance requires sophisticated performance analysis capabilities due to the complexity of execution patterns and underlying hardware, and no tool could handle the voluminous performance trace data needed to detect potential problems. This work introduces Chimbuko, a performance analysis framework that provides real-time, distributed, \textit{in situ} anomaly detection. Data volumes are reduced for human-level processing without losing necessary details. Chimbuko supports online performance monitoring via a visualization module that presents the overall workflow anomaly distribution, call stacks, and timelines. Chimbuko also supports the capture and reduction of performance provenance. To the best of our knowledge, Chimbuko is the first online, distributed, and scalable workflow-level performance trace analysis framework, and we demonstrate the tool's usefulness on Oak Ridge National Laboratory's Summit system.


\end{abstract}

%
%



%


\section{Introduction}
The Chimbuko framework captures, analyzes, and visualizes performance metrics for complex scientific workflows at scale. Meanwhile, the TAU performance analysis tool is used to capture trace and profile performance data, and provenance of the underlying architecture's specification and execution details are extracted. All performance data are analyzed, reduced, and visualized in real time and \textit{in situ} by Chimbuko, providing key insights into the behaviors of scientific applications and workflows. The performance provenance data enable the identification of specific performance events, as well as the comparison of performance behavior across several runs. A key application area for Chimbuko is co-design trade-off studies for online data analysis and reduction workflows because it allows scientific applications and workflows to trace their execution patterns on heterogeneous architectures at realistic scales.


The U.S. Department of Energy (DOE)'s Exascale Computing Project (ECP) currently is developing a portfolio of applications to run on the first exascale systems, and a significant portion of these applications are workflows. Furthermore, scientific workflows are becoming more prevalent in mainstream scientific computing.  
Here, workflows are defined as the composition of numerous coupled tasks. They execute (\textit{in situ}) by exchanging information over memory, storage hierarchy, and interconnect network of a high-performance computing (HPC) system \cite{deelman2018}.

Workflows face specific performance challenges that extend beyond the performance of its separate single components, specifically interdependence between workflow components and increased potential for resource contention. Assessing their performance and identifying possible bottlenecks require tools that exceed today's available performance analysis tools. Along with the ability to capture the interaction between several applications combined within a workflow, the new tools also need to cope with tremendous performance data volumes. Rather than collecting trace-level performance data from a single application, it is necessary to capture it for multiple applications at the same time. Given the cost of data movement and input/output (I/O) on large-scale HPC systems, it is paramount to analyze and reduce this performance data immediately and not introduce unacceptable levels of overhead and distortion to the workflow.
At the same time, critical information from the performance data must be preserved. Therefore, rather than just providing high-level statistics, Chimbuko focuses on anomalous events along with sufficient information for a root cause analysis. Online performance visualization enables workflow developers to investigate anomalies as they occur, for example, to determine if they are workflow internal or caused by contention for resources that other users' applications are heavily using. The identified anomalies also can be saved to disk for more in-depth study later.

The primary characteristic that distinguishes Chimbuko from existing tools is its ability to perform scalable, trace-level, real-time performance analysis on workflows---the first such workflow performance analysis tool available to date. Specifically, Chimbuko provides the following innovative functionality: 
\begin{enumerate}
\item \textbf{Performance Data Analysis across 
Workflows \textit{In Situ}}. In this approach, applications orchestrated as a workflow emit performance anomaly data to a single analysis instance. Thus, we can identify performance bottlenecks stemming from inefficient interactions between workflow components and issues caused by their combined use of system resources in addition to potential bottlenecks within the applications.  

\item \textbf{Data Reduction using Anomaly Detection}. We offer data reduction methods based on anomaly detection (AD) that filter normal events while focusing on abnormal performance behavior as these are of greater interest to application developers.

\item \textbf{Scalable Architecture}. We have developed a novel, distributed architecture for \textit{in situ} performance trace analysis that minimizes data traffic across the communication fabric. We process large quantities of performance trace data on the node where the data are generated. This feature significantly reduces the quantity of inter-node communication and enables real-time analysis. 

\item \textbf{Visualization}. We offer a coupled \textit{in situ} visualization framework to monitor abnormal performance. The visualization presents a multiscale design by dynamically and interactively updating the trace anomalies in different levels of detail, e.g., rank, time frame, function execution, and function call stack. This feature allows the users to monitor the performance online and quickly dive into execution details for root cause investigation. 

\item \textbf{Prescriptive Provenance}. We extract streaming performance provenance information for each run and reduce the trace to focus on detected anomalies. Provenance includes static information for a run and dynamic information, such as workflow execution environment. Metrics are available for collection together with their provenance when anomalies are detected. This supports not only online analysis of anomalies, but also subsequent comparison with other runs.
\end{enumerate}

To show the proposed framework's capabilities, we ran performance studies for several workflows. Among them, a workflow based on the NWChem computational chemistry code \cite{valiev_nwchem_2010} is demonstrated on the Summit supercomputer. We executed the workflow while analyzing trace-level performance data with Chimbuko on thousands of Message Passing Interface (MPI) ranks. Compared to running the workflow with the TAU performance analysis tool \cite{Shende2006} alone, which collects and saves all performance data, we were able to reduce the trace data rate 148 times. The ability to locate performance anomalies quickly and efficiently allowed us to identify communication delay issues in the workflow.

\section{Chimbuko Architecture}

\subsection{Design Objectives}
Our goal is to deliver a real-time, scalable analysis tool for performance trace data that can diagnose workflow-level performance behaviors. The Chimbuko framework's main components are: {\bf{ 1) AD, 2) Visualization, and 3) Provenance}}. The initial aim is to design a system that will scale to: 
\begin{itemize}
    \item \textbf{Nodes:} 1000 or more
    \item \textbf{MPI Ranks:} 10,000 or more.
\end{itemize}

\subsection{Architecture}
\begin{figure*}[tbh]
 \centering
  \includegraphics[width=0.9\textwidth]{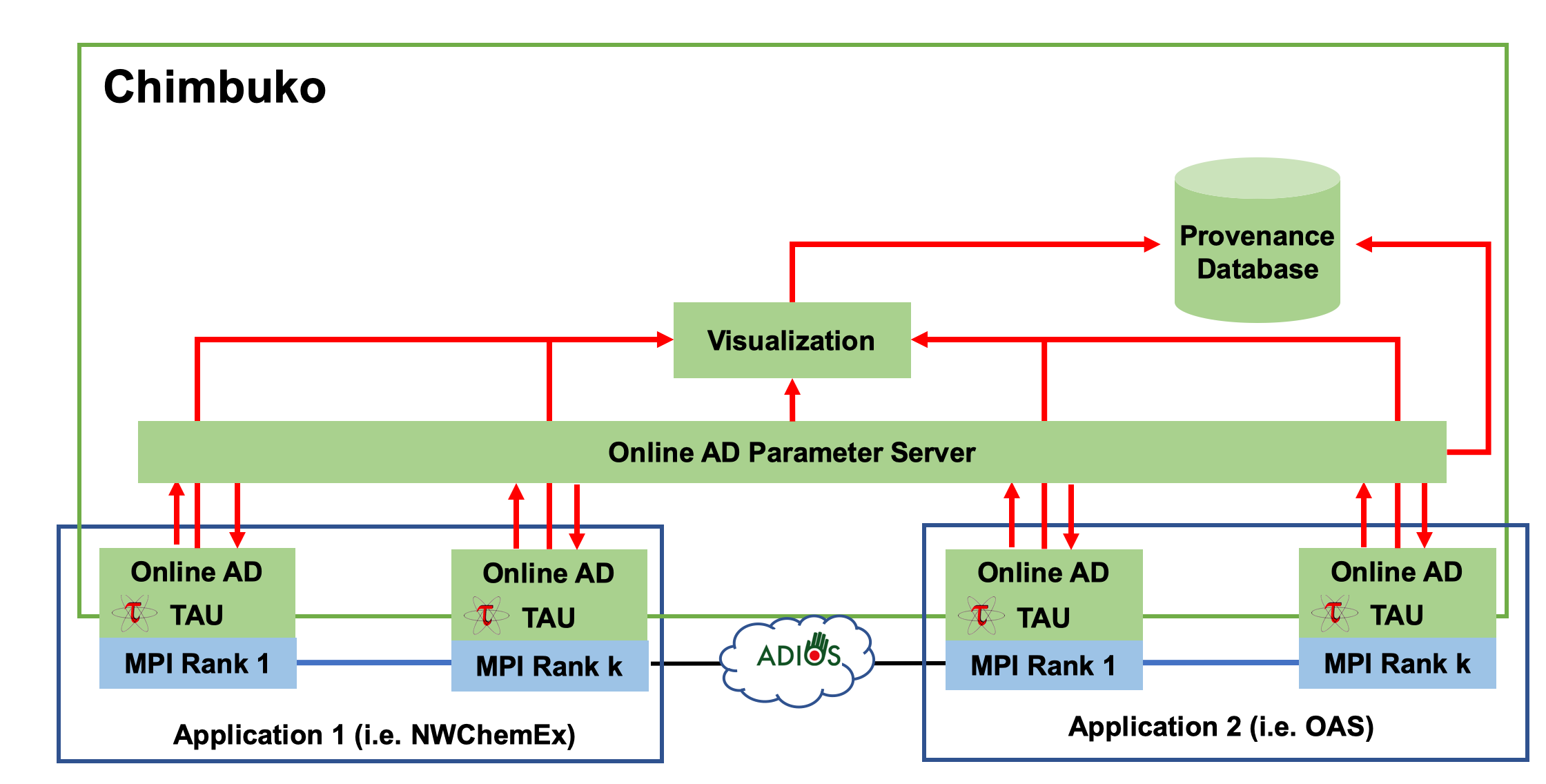}
 \caption{Chimbuko architecture diagram with the major components: TAU, Online AD modules, Visualization, and Provenance Database. Image illustrates two concurrently running applications.}
\label{designfig:2}     
 \end{figure*}
The proposed architecture was designed to enable efficient, distributed data processing; limited data movement; and scalability while generating minimal overheads for the workflow under investigation---all critical characteristics for a tool that needs to operate on future exascale systems. 

Figure~\ref{designfig:2} depicts the proposed Chimbuko architecture diagram, showing a workflow with two applications running concurrently (for simplicity). 
In this architecture, aggregating the performance trace data across the workflow is avoided because relevant events already are selected at each node. The remaining events are used to create performance profile statistics then discarded. Profile data remain useful to give a general impression of the workflow's performance, but it is not supportive of identifying specific performance issues. Online AD is split into two modules: an online AD parameter server and on-node online AD module.


\textbf{Online versus Offline}. The Chimbuko framework has been designed not only to provide support for online performance analysis, but also for optimization efforts that require longer-term performance studies. To this end, all Chimbuko components can be run both in on- and off-line modes, allowing users to reinvestigate and compare performance data across a number of runs. 
The online (\textit{in situ}) components are implemented and available in GitHub repositories~\cite{repo_chim,repo_viz}. The following sections describe the different Chimbuko components in more detail.

\subsection{TAU}
\label{subsec:tau}
The TAU Performance System~\cite{Shende2006} is a portable profiling and tracing toolkit for performance analysis of parallel programs written in Fortran, C/C++, Java, and Python. TAU is capable of gathering performance information through system-interrupt-based sampling and/or instrumentation of functions, methods, basic blocks, and statements. The instrumentation can be inserted in the source code automatically with a TAU-specific compiler wrapper based on the Program Database Toolkit (PDT)~\cite{Lindlan2000}, dynamically using DyninstAPI~\cite{williams2016dyninst}, at runtime in the Java Virtual Machine or Python runtime, or manually using the instrumentation API (application programming interface). TAU measurements represent first-person, per-OS (operating system) thread measurements for all processes in a distributed application, such as an MPI simulation. TAU measurements are collected as profile summaries and/or a full event trace.

To enable runtime analysis of TAU performance data, TAU has been extended with an ADIOS2 plugin. ADIOS2 is a reimplementation of the \textit{ADaptable Input Output System}, ADIOS~\cite{Lofstead2008}. In addition to file support, ADIOS provides a step-based (\textit{in situ}) data engine that can be read by external applications running concurrently with a scientific simulation. Using the available ADIOS2 Sustainable Staging Transport (SST) engine, TAU trace data periodically are written to an output data stream to be consumed by the Chimbuko analysis.

In the experiments described herein, NWChem was compiled using TAU compiler wrappers, selectively instrumenting to avoid high-frequency, short-duration functions. TAU also includes an MPI interposition library to measure MPI calls without requiring instrumentation. As instrumented functions enter and exit and when send/receive events happen, events are written to a local buffer. Periodically, the event buffer is written to the ADIOS2 data stream. The periodicity for the following experiments was configured for once-per-second as a compromise between providing fresh data and minimizing overhead.

\section{Performance Anomaly Detection}

For modern scientific workflows on exascale systems, the trace data generated by a representative workflow would amount to hundreds of terabytes of data---most of which are uninteresting from a developers' perspective, except for events that could indicate potential performance bottlenecks. For example, the execution time of a parallel program is bounded by the slowest processor. Therefore, parallel application developers aim to distribute the computational work evenly among all processors, but this may not always be achieved. If in a problem setting operations take disproportionately more time on one processor than on others, this indicates the processor was assigned more work, pointing to a potential bottleneck. Analyzing program trace data helps to detect these anomalies. However, this analysis cannot happen postmortem without storing prohibitive amounts of data on disk. To date, application developers either had to run much smaller problem sizes, potentially not capturing the behavior they wanted to investigate, or only look at very small areas of code at any given point in time, possibly missing the root cause or the event itself completely. We propose that the analysis of the full trace data should be performed on the fly, providing access to all of the necessary information while quickly reducing the collected data to only their significant components.  

\subsection{Setup}
As mentioned previously, TAU is used to instrument workflow application source codes and generate trace data. During execution of the Chimbuko workflow, TAU submits the observed events to our analysis code, which performs the data reduction. There are largely two types of events: \textit{function} and \textit{communication} events. A function event contains information of the function identifier, name, and its type (ENTRY or EXIT). A communication event includes data tag and size (in bytes) and identifiers of sender and receiver. All events come with common information, including identifiers of the application, MPI rank and thread, and timestamp (in micro-seconds). 

\subsection{Data Reduction and Anomaly Detection} \label{sec:ad_module}
This section introduces Chimbuko's methodology, used to reduce performance data and  detect anomalies, for example, within the function execution times. As execution time imbalances are a major source of workflow performance variability, we start with this particular metric. It is our intention to add the analysis of further metrics to Chimbuko in upcoming releases. The major components are the \textbf{On-node AD Module} and \textbf{Parameter Server}.

\subsubsection{On-node AD Module} 
The on-node AD module takes streamed trace data (per rank) from TAU. As all events in the streamed trace data are sorted according to the event timestamp, the AD module can build and maintain a function call stack with function events and map communication events to a specific function if they are available. The completed function calls (i.e., ENTRY and EXIT of a function event are observed) are extracted from the tree and statistical analysis based on the function execution time is performed to determine anomalous function calls that have extraordinary execution times compared to normal function calls. Specifically, we label a function call as an anomaly if it has a longer execution time than the upper threshold ($= \mu_{i} + \alpha * \sigma_{i}$) or has a shorter execution time than the lower threshold ($= \mu_{i} - \alpha * \sigma_{i}$), where $\mu_{i}$ and $\sigma_{i}$ are mean and standard deviation of execution time of a function $i$, respectively, and $\alpha$ is a control parameter that is set to $6$ in our entire studies. 

If there are any anomalies within the current trace data, the AD module sends them to visualization server or stores them in files. This is where significant data reduction occurs because we only save the anomalies and a few nearby normal function calls of the anomalies. In our studies, we saved anomalies along with most $k$ normal functions calls before and after the anomaly (if available). For our case studies, we set $k$ to be $5$. 

To have consistent and robust AD power, it is important to update local statistics ($\mu_{i}$ and $\sigma_{i}$) in each on-node AD module with the global one that is available in the Parameter Server. Along with local statistics, the AD module also sends other information, such as the number of detected anomalies to the Parameter Server, so the server can have a global view of workflow-level performance trace analysis.


\subsubsection{Online AD Parameter Server}
The online AD Parameter Server is designed to maintain a global view of workflow-level performance trace analysis. It includes, for example, execution time statistics per function ($\mu_{i}$ and $\sigma_{i}$) and the number of detected anomalies per rank within a time interval, which is determined by the frequency of TAU streaming data (refer to Section~\ref{subsec:tau}). The global view is updated by the local analysis results from the on-node AD modules without any synchronization barriers and periodically sent to the visualization server if there are any updates. This enables users to review the performance trace analysis results overview in real time. In our case studies, the periodicity is set to $1$ second.

\section{Performance Visualization}
The design goal for visualization is to provide an online anomaly visualization and exploration platform that not only allows users to observe any anomalous performance, but also to support a deeper investigation of specific anomalies. This platform must cope with significant data streams. Therefore, migrating from our previous versions~\cite{Xie2018}\cite{xie2019}, we divide the visualization into two parts: \textit{in situ} anomaly statistics visualization and online analysis visualization. In the following, we first describe the visualization backend server architecture followed by a detailed explanation of the two visualization parts.

\subsection{Visualization Backend Server}

\begin{figure}[!tb]
\centerline{\includegraphics[width=0.5\textwidth]{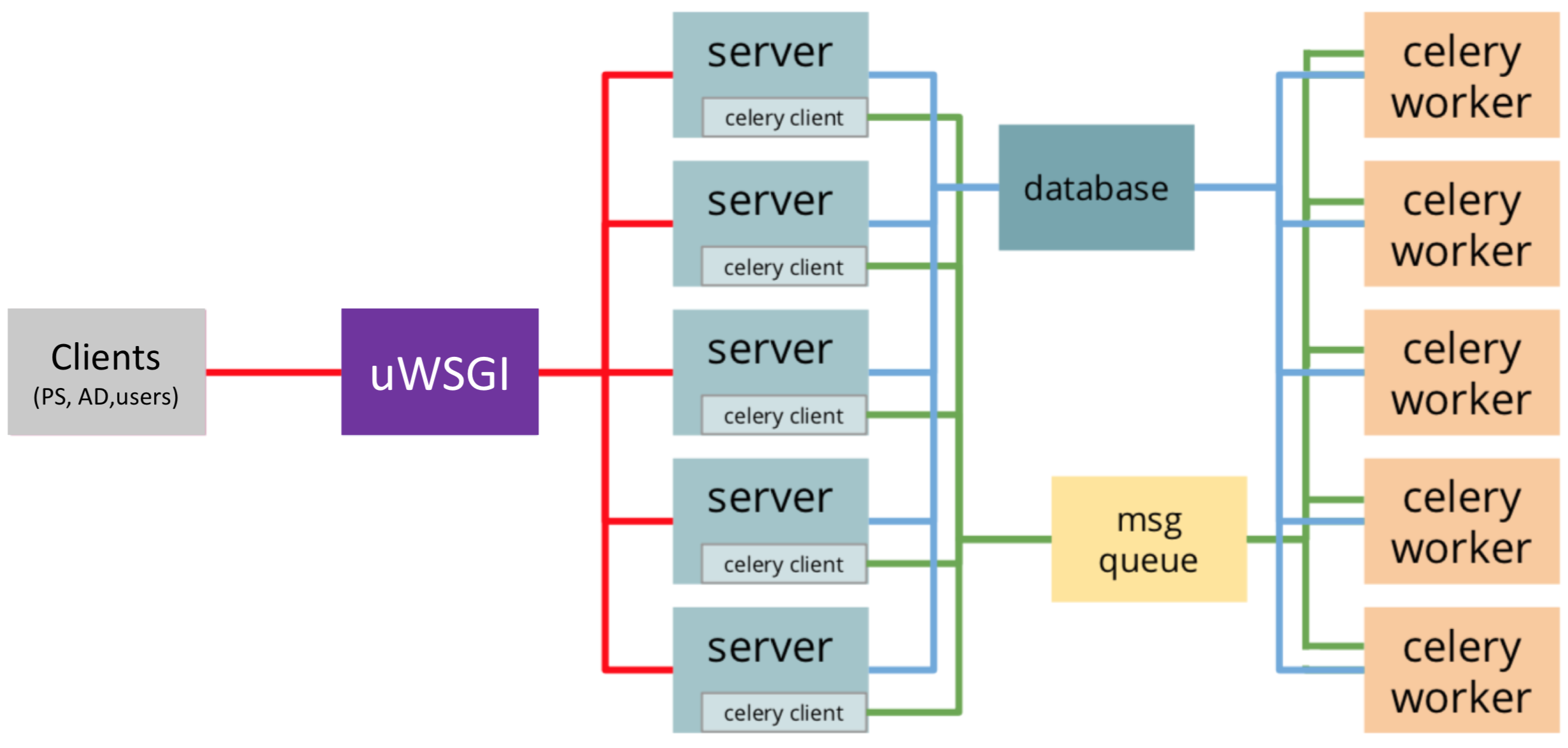}}
\caption {The architecture for visualization backend server \cite{flaskAtScale} adopted for this work. Specifically, it includes uWSGI workers serving concurrent connections, celery workers and a Redis message queue for asynchronous tasks, and a database for data storage. }
\label{fig:vizserver}
\end{figure}

There are two types of clients of the visualization server. The first is data senders, including the on-node AD modules and the online AD Parameter Server. The second type is the users who are actually exploring and interacting with the proposed visualization via a web browser. 

For the data senders, the visualization server should be able to digest the requests asynchronously so there is no waiting time and minimal memory overhead on the senders' side by consuming data as quickly as possible. Furthermore, the visualization server should be able to handle huge amounts of concurrent requests---as many as the number of on-node AD modules currently running. For users, the visualization server should be able to stream data as soon as they arrive from the data senders (e.g., Parameter Server). In addition, it must manage long-running tasks asynchronously, so users can interact with the visualization without it freezing as data are updated. 

To meet these requirements, the visualization server is designed to have two levels of scaling \cite{flaskAtScale}. At the first level, uWSGI \cite{uwsgi} will instantiate the web application in its first process and will fork multiple times until the desired number of workers is reached. Each of them will be fully instantiated to be ready to serve connections. At the second level, especially to handle the long-running tasks (e.g., inserting/querying data into/from database) asynchronously, the requests are distributed over pre-forked processes (celery workers \cite{celery} and Redis message queue \cite{redis}) and return responses as quickly as possible. Finally, streaming (or broadcasting) data to the connected users is done by using Websocket technology with socket IO library. Figure \ref{fig:vizserver} depicts the overall architecture of the visualization server.

In its current implementation, we use a simple, serverless SQL database (SQLite) that has a limitation on handling concurrent writing operations. Thus, the data from the on-node AD modules are stored in predefined file paths directly by the on-node AD modules, and the visualization server fetches those files upon users' requests.

\subsubsection{Requests from Data Senders} All requests from the data senders are processed asynchronously and stored in the SQL-based database. From this, the data from the Parameter Server are broadcast to the connected users after processing the data according to users' query conditions. The broadcast data are used for the \textit{in situ} visualization.

\subsubsection{Requests from Users} Requests from users are processed synchronously for simple tasks (e.g., change query conditions) and asynchronously for long-running tasks (e.g., query to database). One example of long-running tasks is to query a call stack that contains anomaly functions within a certain time range and a specific MPI process of an application. The queried data are broadcast to the connected users using socket I/O.

In the next two subsections, two frontend visualization modules are introduced. This visualization design presents data in different levels of detail by following the ``Overview first, zoom and filter then details on-demand" mechanism commonly adopted in the visualization domain \cite{card1999readings}. 

\subsection{\textit{In Situ} Visualization}
\begin{figure}[!tb]
\centerline{\includegraphics[width=0.4\textwidth]{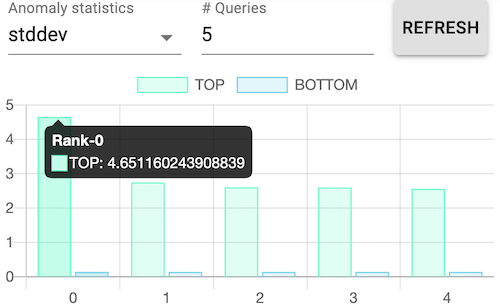}}
\caption {Dynamic ranking dashboard as rank-level visualization granularity for the most and least problematic MPI processes.}
\label{fig:anomaly_stat_view}
\end{figure}

\begin{figure}[!tb]
\centerline{\includegraphics[width=0.5\textwidth]{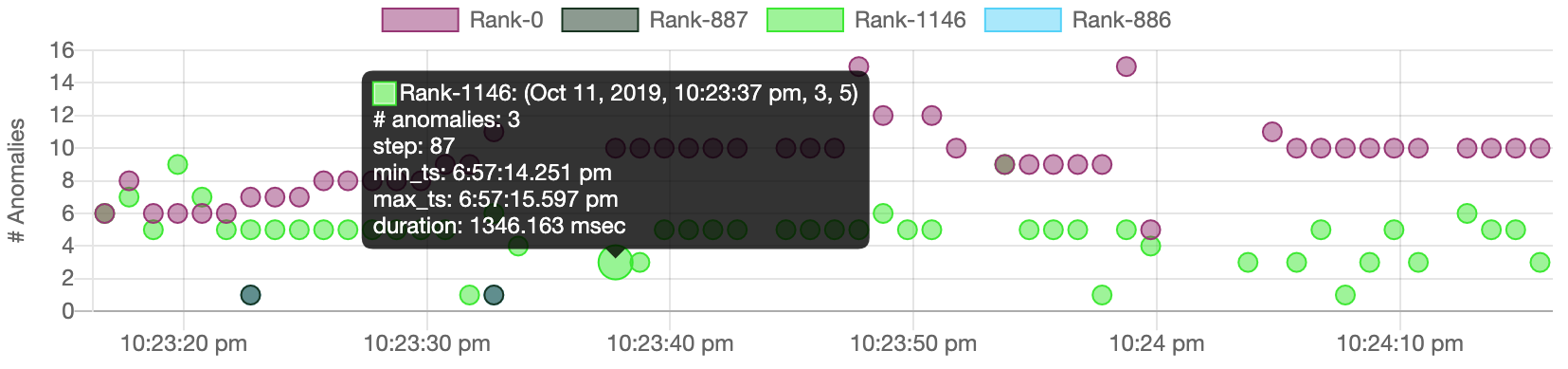}}
\caption {Streaming view of several series of the numbers of anomalies as time-frame-level visualization granularity for various ranks.}
\label{fig:anomaly_history}
\end{figure}

Using data from the Parameter Server, the \textit{in situ} visualization receives data in a streaming fashion and processes it into a number of anomaly statistics. We aim to provide a dynamic ``ranking dashboard" of the most and least problematic MPI ranks as a rank-level granularity of the application. Figure \ref{fig:anomaly_stat_view} shows the statistics per rank (or MPI processes) adopted to select the focused ranks. The statistics includes the average, standard deviation, maximum, minimum, and total number of anomaly functions. Users can select one of the statistics along with the number ranks they want to see. For example, Fig. \ref{fig:anomaly_stat_view} shows the top and bottom $5$ ranks based on standard deviation. Detailed information is available when hovering over each bar chart. 

Selecting (or clicking) bars in Fig. \ref{fig:anomaly_stat_view} activates the visualization server to broadcast the number of anomalies to the connected users in a time frame while performance traced applications are running. This streaming scatter plot serves as a time-frame-level granularity by showing the number of anomalies of an MPI rank within a time window. In Fig. \ref{fig:anomaly_history}, each dot represents a time frame (referred to as ``step'' in the case study) of a selected rank. Color encoding is applied to differentiate ranks. Hovering over a dot will provide a pop-up window that shows detailed information, including the number of detected anomalies, time frame identification (ID), and analyzed time range. Clicking one dot triggers the online analysis visualization.

\subsection{Online Analysis Visualization}

\begin{figure}[!b]
\centerline{\includegraphics[width=0.4\textwidth]{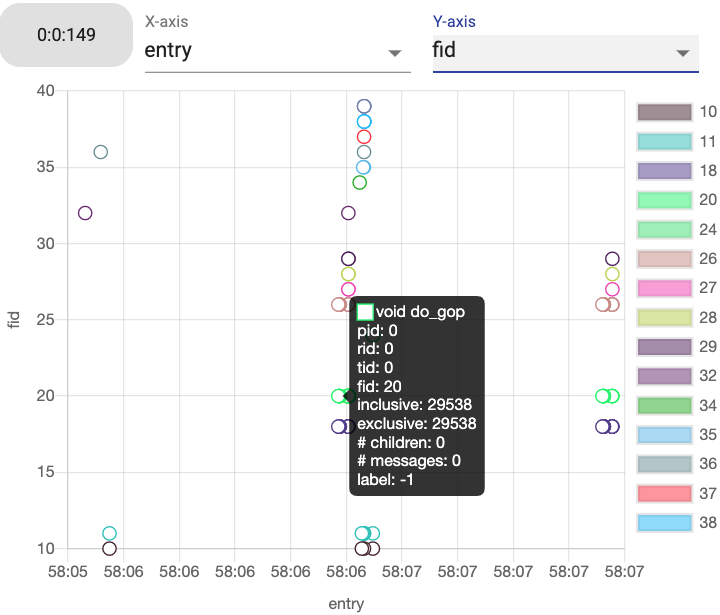}}
\caption {The function execution visualization for a selected time frame (ID 149) and Rank ID 0 for Application 0 in an ``entry" versus ``fid" layout.}
\label{fig:anomaly_func_view}
\end{figure}

The online analysis visualization is designed to retrieve data from a database and present the finest level of granularity into the function execution details. It consists of two parts: a \textit{function} view and \textit{call stack} view. The function view visualizes distribution of executed functions within a selected time interval (Fig. \ref{fig:anomaly_history}). The distribution can be controlled by selecting the X- and Y-axis among function ID, entry and exit time, inclusive and exclusive runtime, label indicating if it is anomaly or not, number of children functions, and number of communication (messages). For example, Fig. \ref{fig:anomaly_func_view} shows some of executed functions at $0$-th application, $0$-th rank, and $149$-th frame. For the X- and Y-axis, entry time and function ID are selected, respectively. Hovering over a circle will show all available information. Clicking a circle (or a function) will trigger a call stack view that includes the selected function.

\begin{figure}[!tb]
\centerline{\includegraphics[width=0.5\textwidth]{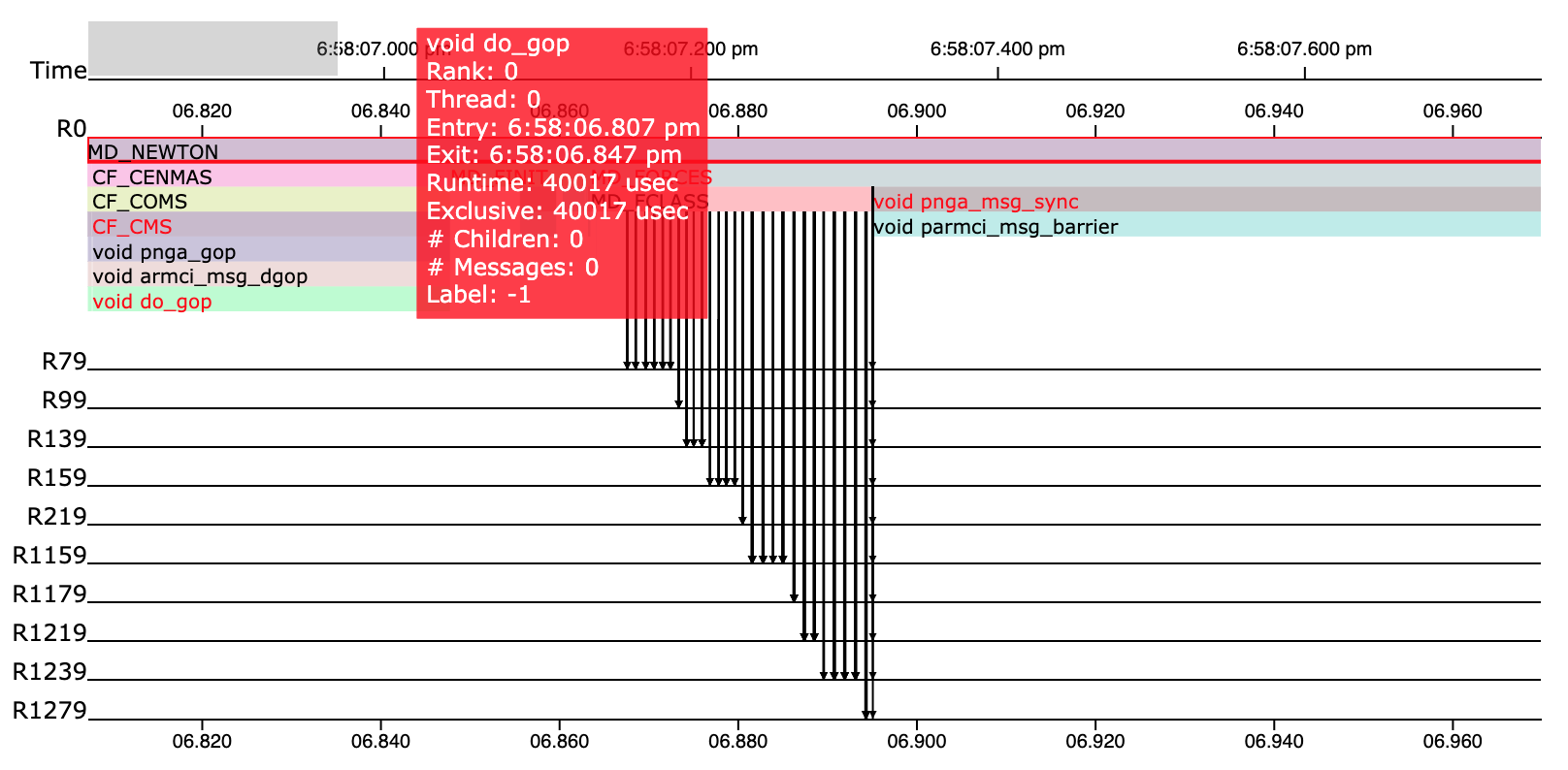}}
\caption {The call stack visualization for the selected function ``MD\_NEWTON."}
\label{fig:anomaly_call_stack}
\end{figure}

In the call stack view, users can more closely investigate the relationships among functions and communications over ranks. For example, Fig. \ref{fig:anomaly_call_stack} shows a zoomed-in call stack view within the time range refined by a user. Anomaly labels are encoded by the color of the name of each function with black being normal and red indicating abnormal. Hovering over a horizontal bar in the call stack will provide a pop-up window with detailed information regarding the corresponding function. Communication (message receiving or sending) is encoded by an arrow between a function and a horizontal line representing another rank ID.

\section{Prescriptive Provenance}
We have developed the concept of \textit{prescriptive provenance} to facilitate the extraction of provenance for selected performance events, describing their execution stack and environment. Prescriptive provenance is the provenance of events identified as anomalies by the distributed AD. The AD prescribes the events for which provenance is extracted. In addition, prescriptive provenance includes a number of functions executing before and after these events are collected.  Some provenance data are collected directly by the TAU tools, such as static information for a run, architecture and software libraries, TAU instrumentation variables, and filtering configuration used. Other information is generated by Chimbuko from the performance trace, such as the anomalous functions and their rank, thread, entry and exit timestamps, runtime, number of children and messages, and a label (normal or anomaly).

Anomalous events are stored on disk together with their provenance by the On-node AD Module. Additionally, a user-defined runtime variable indicates the number of functions to be stored before and after a detected anomaly. The value determined by heuristics depends on the application code and its stability regarding performance. 

Provenance metadata will provide a map of the execution model for each workflow run, including dataflow patterns. Prescriptive provenance takes advantage of the data reduction afforded by AD to reduce a verbose performance trace while simultaneously exhibiting anomalous and surrounding functions at a detailed level.

In future work, when a locally anomalous event is detected, the On-Node AD Module will output provenance data and metadata. In the case of a globally detected event, the online AD Parameter Server will trigger the provenance data and metadata output for all nodes involved.

\section{Experiments}
In this work, we have selected a performance study that was executed on Summit~\cite{vazhkudai_design_2018}, a supercomputer equipped with IBM POWER9 CPUs (2/node) and Volta V100s GPUs (6/node) at the Oak Ridge Leadership Computing Facility (OLCF). Each node is furnished with 512 GB of DDR4, 96 GB HBM2, and 1600 GB of NV Memory. Summit nodes are connected by a Mellanox EDR 100G InfiniBand network with non-blocking fat tree topology.

\subsection{The NWChemEx Scientific Use Case}
The NWChemEx ECP project is targeting a range of computational chemistry methods, from molecular dynamics to high-order many-body methods. For molecular dynamics capabilities, NWChemEx steers toward simulations of about one million atoms, simulating processes on timescales of about a microsecond and taking about one billion time steps to complete \cite{NWChemEX_Case}. Each time step creates a snapshot of the molecular structure, and the time sequence of these shots forms a trajectory. To cope with the resulting data volume generated by these simulations, trajectories will need to be analyzed on the fly. This approach requires a workflow setup where the simulation and analysis codes run concurrently and the data flow from one component to another. From a performance perspective, the workflow approach generates a number of additional concerns related to the scalability of different components, the data flows between them, and how interactions between the components affect performance. 
The NWChemEx project's goal is to achieve high performance, requiring nontrivial optimization efforts, and its effectiveness must be measurable to demonstrate any benefits. Application performance provenance information (i.e., knowledge about the development of code performance over time) can help identify performance bottlenecks faster, as well as document the effectiveness of performance optimization efforts~\cite{pouchard_capturing_2017}. Helpful information types include provenance data regarding how the workflow maps to the machine (thereby defining dataflow patterns) and where in the calculation (processor, function, thread, etc.) performance issues arise. It is expected that performance issues often will be related to communication. This can be within a node if busses get saturated or between nodes if the network gets saturated or load imbalance forces processes to wait. Thus, it will be important to record where data originate and where it flows. Maintaining a historical record of this information also is useful, so multiple simulations can be run with different workflow configurations in a co-design study. Moreover, this information can be mined to discover how anomalous patterns depend on the workflow configuration. The ability to query the execution of a single workflow and analyze performance anomalies that arise will assist in identifying problems and suggesting solutions. Meanwhile, the ability to query how performance changes across different simulation setups will help assess what has been achieved and how much of the design space has been explored. 

As NWChemEx is under development, a modified version of the molecular dynamics capability in NWChem~\cite{valiev_nwchem_2010} was used for the case study provided in this work. The biomolecular system considered in this case represents the scale and complexity of the systems that NWChemEx ultimately will be able to simulate on next-generation leadership-class computers. The molecular system consists of two lipid layers in an aqueous environment, while transmembrane proteins are embedded in the lipid layers~\cite{vandam2019}. Using periodic boundary conditions, this system emulates cellular compartments that can interchange calcium ions through transmembrane ion channels. The system consists of 1.2 million atoms to run simulations at a realistic scale. To demonstrate \textit{in situ} data analysis, NWChem was modified to stream the trajectory data through ADIOS2 to the analysis component. In addition, NWChem and the analysis component were instrumented with TAU to stream performance trace data to the Chimbuko infrastructure and analyze the workflow behavior.

\subsection{Scalability Analysis}\label{sec:exp_scale}
\subsubsection{Analysis of On-node AD Module}
\begin{figure}[!tb]
\centerline{\includegraphics[width=0.5\textwidth]{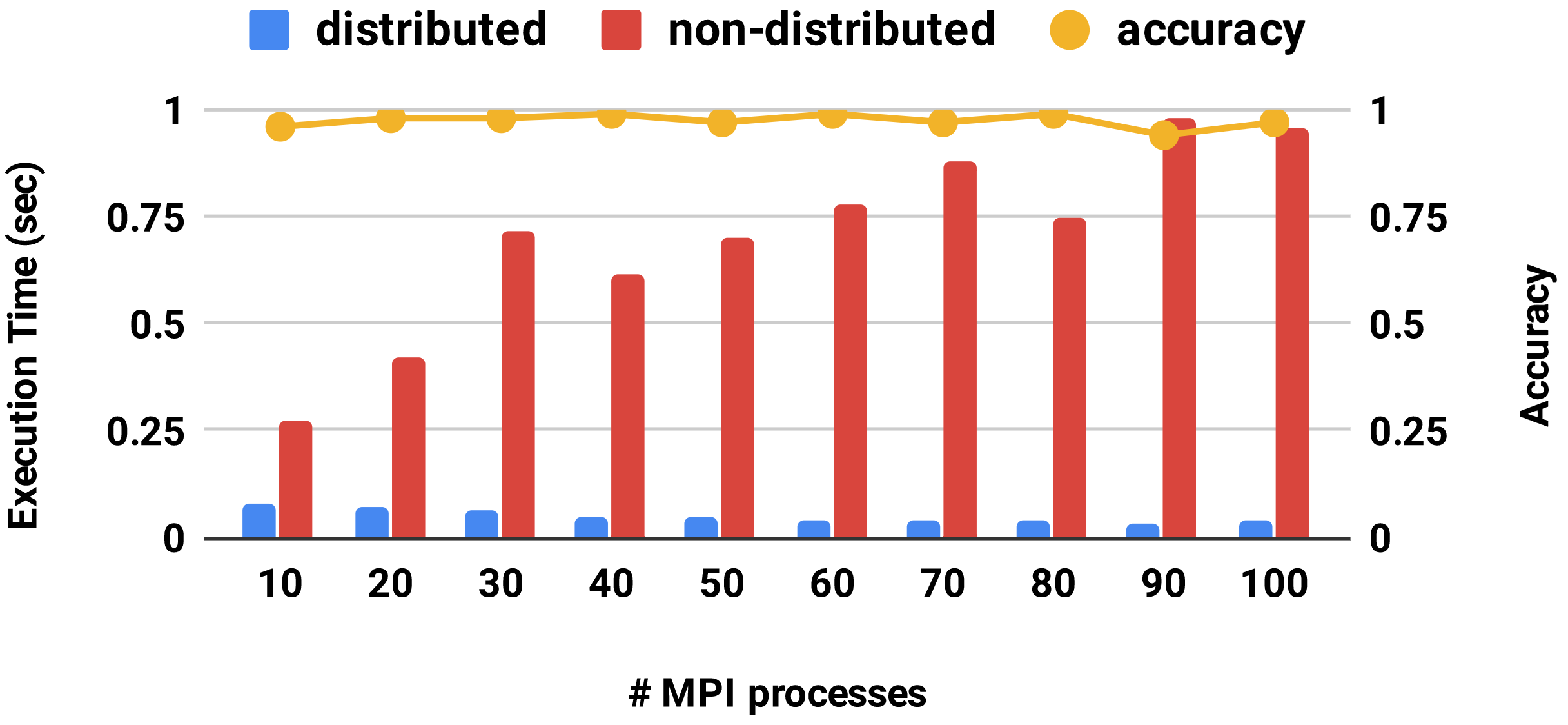}}
\caption {Comparison of the distributed and non-distributed AD modules.}
\label{fig:ad_accuracy}
\end{figure}

The AD accuracy of the on-node AD Module is studied by comparing the distributed and non-distributed versions. In the non-distributed version, all performance data from all MPI processes are managed with a single AD Module instance. As a result, this single instance has the exact statistics. In the distributed version, each MPI process has its own AD Module instance, which handles only the performance data of the corresponding MPI process of the simulation. Therefore, it only has local statistics. These statistics are communicated asynchronously with the Parameter Server, which collects global statistics. The Parameter Server provides the global statistics back to the various AD Module instances by request. The AD Module instances subsequently use a combination of local and global statistics to detect anomalous function behavior. Throughout the simulations, the statistics are updated according to~\cite{pebay2008formulas}. Figure~\ref{fig:ad_accuracy} compares the function calls that were considered anomalous between the non-distributed and distributed versions of the AD Modules. In our experiment, the distributed version is as accurate as the non-distributed version---$97.6\%$ accuracy on average over a range of MPI processes from $10$ to $100$. More importantly, the distributed version shows significantly faster and constant execution time over increasing MPI processes than the non-distributed version. For the distributed version, the average execution time over different numbers of MPI processes is around $0.05$ seconds, while the non-distributed version tends to increase the execution time as the number of processes increases. This is because the single AD instance has to aggregate and process trace data from all MPI processes.

\subsubsection{Analysis of the NWChem Execution Time and Data Reduction}

\begin{figure}[!b]
\centerline{\includegraphics[width=0.5\textwidth]{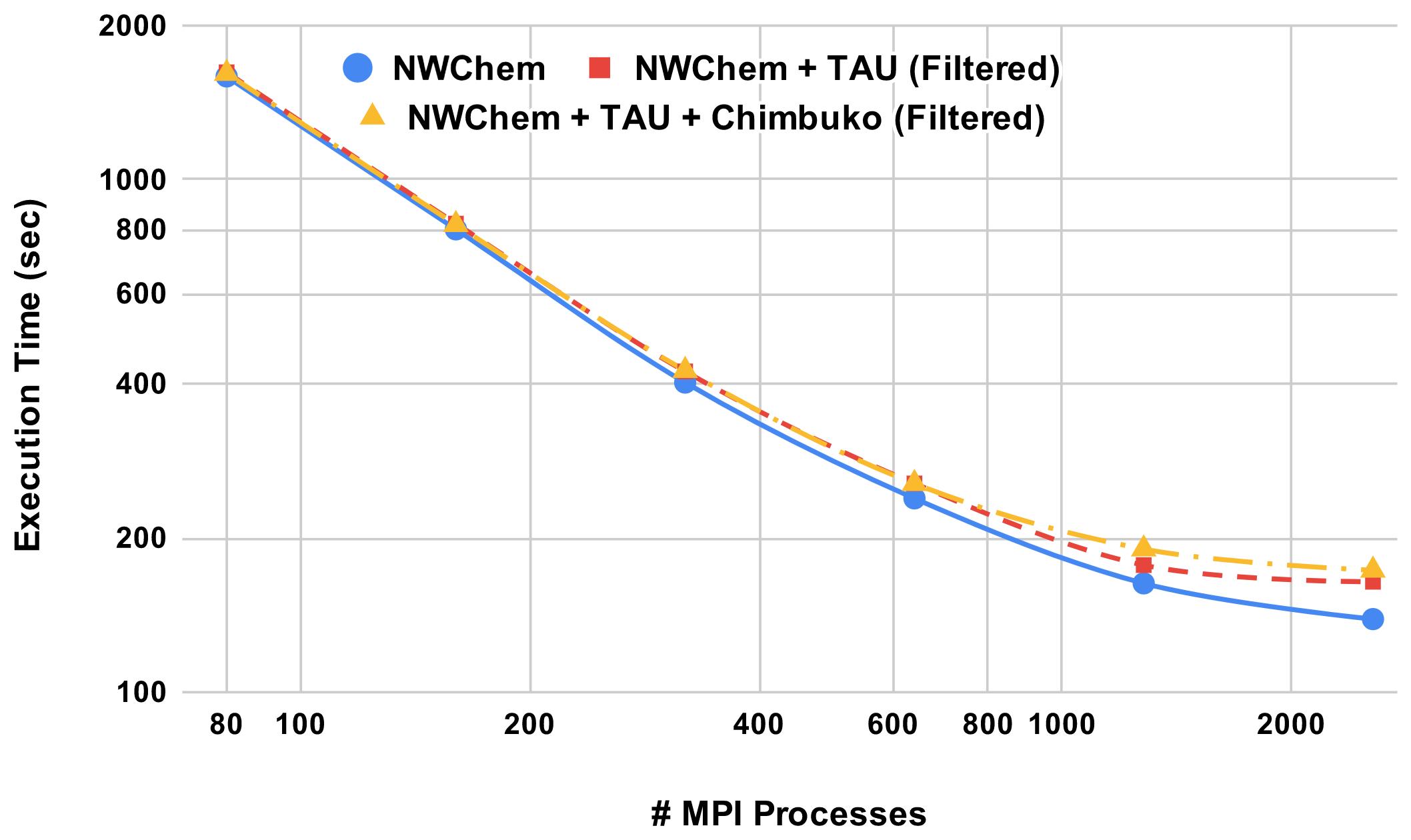}}
\caption {NWChem execution time over MPI processes (log-log).}
\label{fig:nwchem_exectime}
\end{figure}

\begin{figure}[!b]
\centerline{\includegraphics[width=0.5\textwidth]{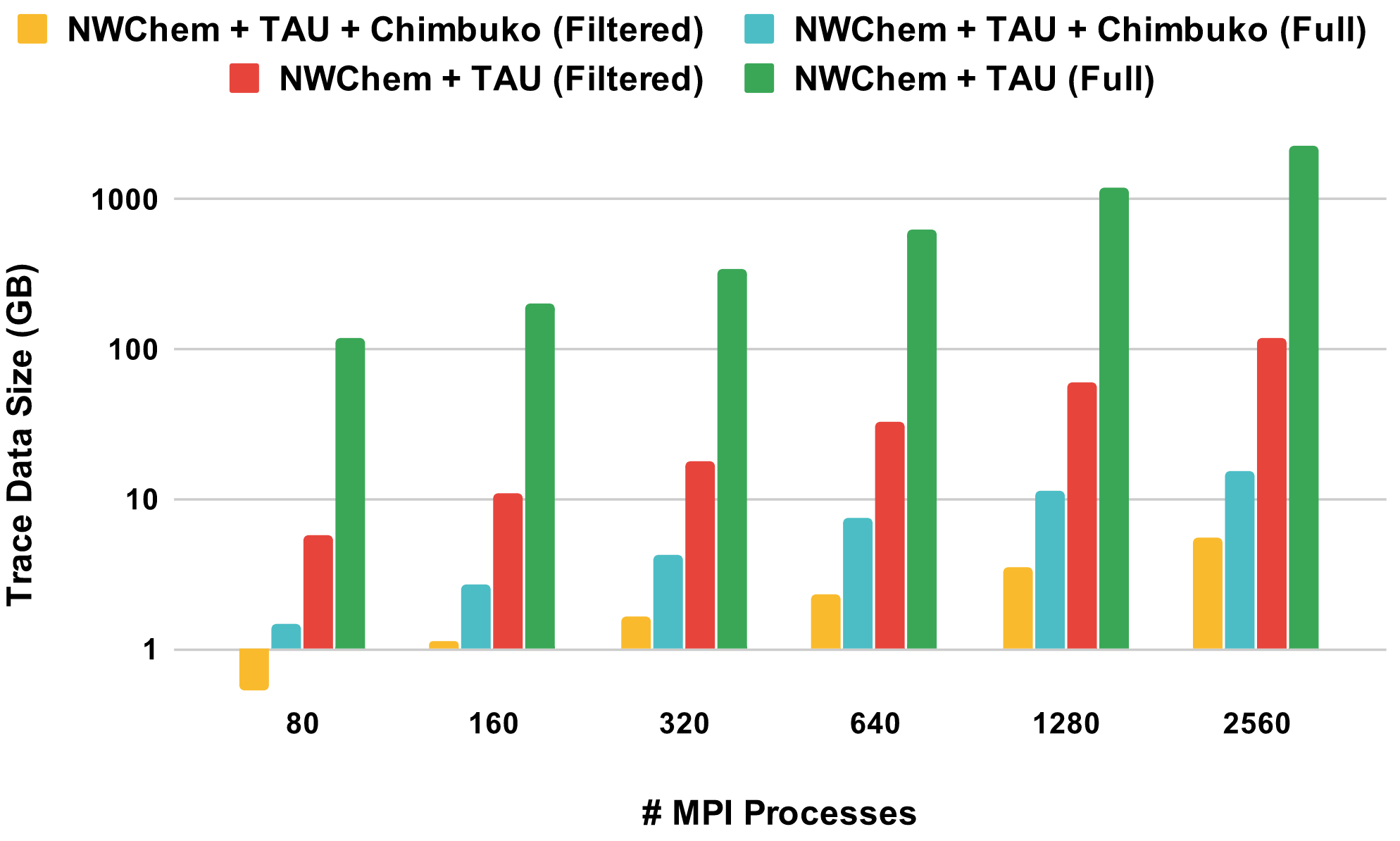}}
\caption{Trace data size over MPI processes.}
\label{fig:data_reduction}
\end{figure}

Figure~\ref{fig:nwchem_exectime} illustrates the execution time of the NWChem workflow measured for three different cases: 1) NWChem, 2) NWChem + TAU, and 3) NWChem + TAU + Chimbuko. Notably, each measurement is the average of $15$ independent executions, while all runs used the same problem size. In addition to measuring the execution time, TAU trace data are collected as NWChem is running. For the "NWChem + TAU" case, the TAU data are dumped into BP files with the ADIOS2 BP engine. For the "NWChem + TAU + Chimbuko" case, the TAU data are streamed to Chimbuko with the ADIOS2 SST engine, and data reduced by Chimbuko are dumped into JSON files. Figure~\ref{fig:data_reduction} measures and compares the sizes of the dumped data. It is worth noting that with help from NWChem team domain scientists, we filtered TAU trace function events in NWChem compilation and running time that do not help the performance bottleneck reasoning. For the execution time, we only used the filtered NWChem + TAU (+ Chimbuko) version. Meanwhile, for the data reduction, we also measured the size of dumped data from the unfiltered version.

Figure~\ref{fig:nwchem_exectime} shows there are not many execution time differences with TAU or TAU + Chimbuko until around $1000$ MPI processes, compared to NWChem only case. After $1000$ MPI processes, the time difference increases. The time difference is reinterpreted as execution time overhead, defined as follows:
\begin{equation}
    overhead (\%) = \frac{(T_{r} - T_{r}^{m})}{T_{r}} \times 100 .
\end{equation}
Here, $T_{r}$ is NWChem ``only'' execution time with $r$ MPI processes, and $T_{r}^{m}$ is NWChem + $m$ execution time, where $m$ is either TAU (without Chimbuko) or TAU + Chimbuko (with Chimbuko). Table~\ref{tb:overhead} summarizes the calculated overhead. With less than $1000$ MPI processes, the overhead from ``with Chimbuko'' is less than $10\%$ (or less than $1\%$ add from ``without Chimbuko''). With larger than $1000$ MPI processes, the overhead suddenly jumps to a higher number. For instance, with $1280$ MPI processes, there is about $8.54\%$ overhead without Chimbuko and about $16.67\%$ overhead with Chimbuko (about $8\%$ add from ``without Chimbuko''). We currently are investigating where the sudden overhead jump comes from. Nevertheless, with the insignificantly increased overhead from Chimbuko (maximum about $8\%$ addition at $1280$ MPI processes) in the execution time, we achieved averages of $14$ and $95$ times of data reduction for filtered and unfiltered (full) cases, respectively. With the largest MPI processes, we achieved $21$ and $148$ times of data reduction, respectively. For example, at $2569$ MPI ranks, $2,300$ GB of raw data (or $117.5$ GB for filtered) are reduced to $15.5$ GB (or $5.5$ GB for filtered)---a factor of $148$ (or $14$) times data reduction with about $6\%$ of additional overhead on top of about $18\%$ overheads from ``without Chimbuko.'' More importantly, even with such a huge reduction in performance data, Chimbuko still can provide scientists/developers  with important insights into workflow performance (demonstrated in Section \ref{sec:case_study}).

\begin{table}[!t]
\caption{Chimbuko overhead over NWChem execution time}
\begin{center}
\begin{tabular}{ c|c|c|c|c|c|c } 
\# MPI & 80 & 160 & 320 & 640 & 1280 & 2560 \\
\hline
without Chimbuko & 1.85 & 2.60 & 5.13 & 6.92 & 8.54 & 18.27 \\
with   Chimbuko & 1.31 & 2.13 & 5.53 & 6.85 & 16.67 & 24.56
\end{tabular}
\end{center}
\label{tb:overhead}
\end{table}

\subsection{A Visual Analysis Case Study}\label{sec:case_study}

\begin{figure}[!t]
\centerline{\includegraphics[width=0.5\textwidth]{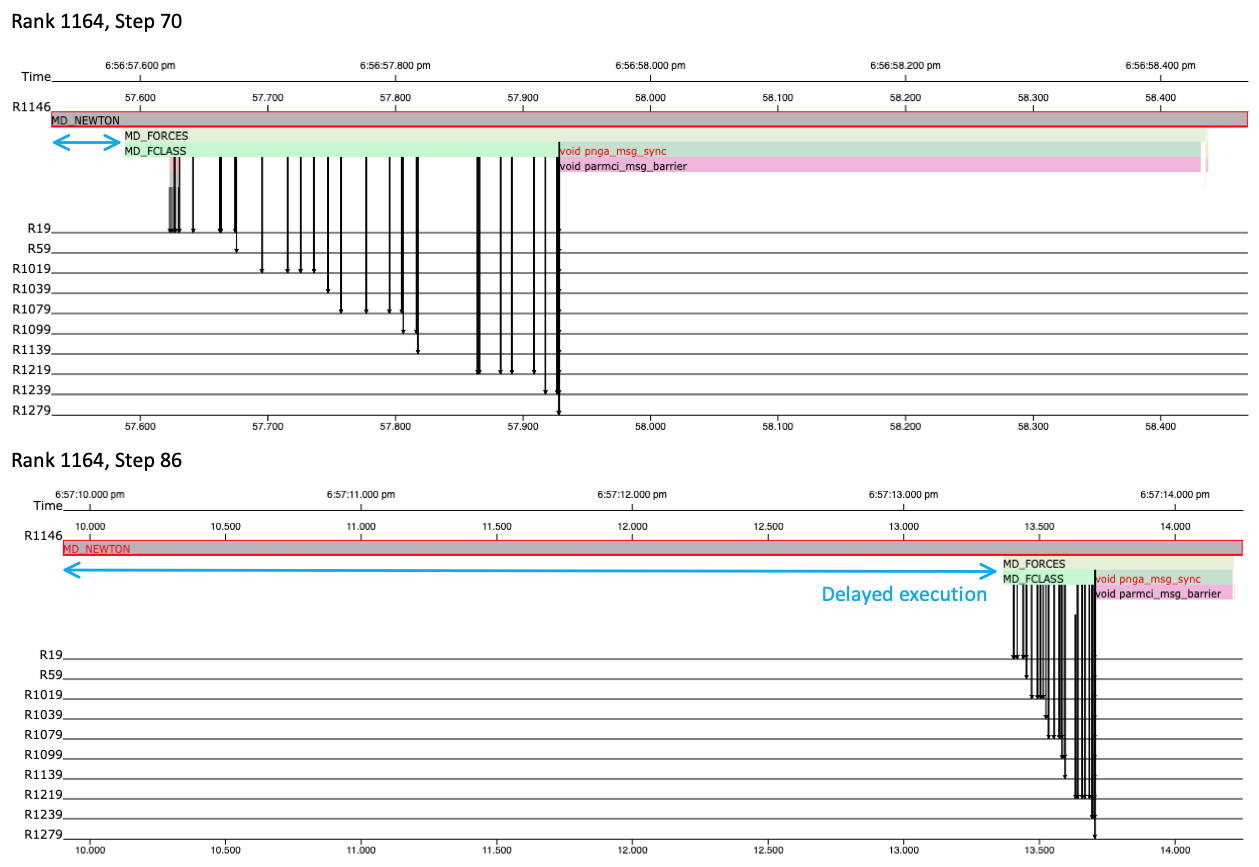}}
\caption{A case study of NWChemEx showing the function delay.}
\label{fig:md_newton}
\end{figure}

We conducted a visual analysis of the performance data with a domain scientist from the NWChemEx team. 

In the experiment, the scientist was specifically interested in the function ``MD\_NEWTON" as it is a major simulation function. By choosing one of the top five anomalous ranks, he selected Rank 1164 and kept tracking the dynamic scatter plot for the step-wise anomaly status. He found that a consecutive step series reported normal execution, such as step 70 shown in the top of Fig.~\ref{fig:md_newton}. However, one execution was identified as an anomaly in step 86, shown in the bottom of the same figure. The abnormal execution almost tripled the time of the normal one. By carefully comparing the children functions in both steps, they remained quite similar. The scientist confirmed that the root cause was not from the children. As highlighted in the figure, he found that the launch time of a children function ``MD\_FORCES" had an apparent delay in the abnormal case. Hence, the scientist concluded the delay as the major reason for decreased performance.

\begin{figure}[!b]
\centerline{\includegraphics[width=0.5\textwidth]{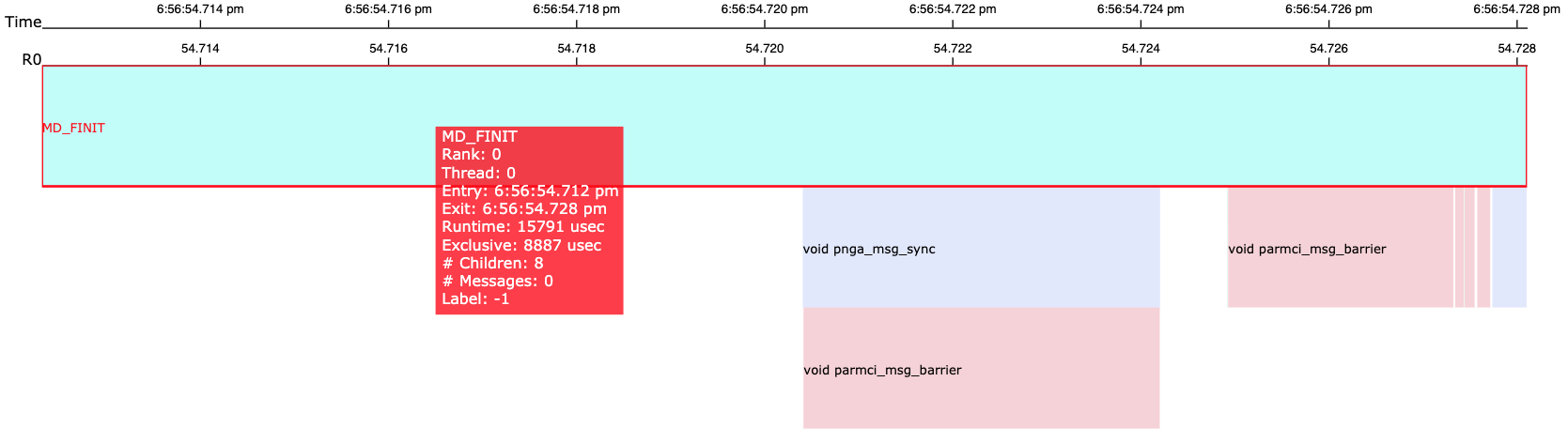}}
\caption{The case study showing anomalous function ``MD\_FINIT."}
\label{fig:md_finit}
\end{figure}

\begin{figure}[!t]
\centerline{\includegraphics[width=0.5\textwidth]{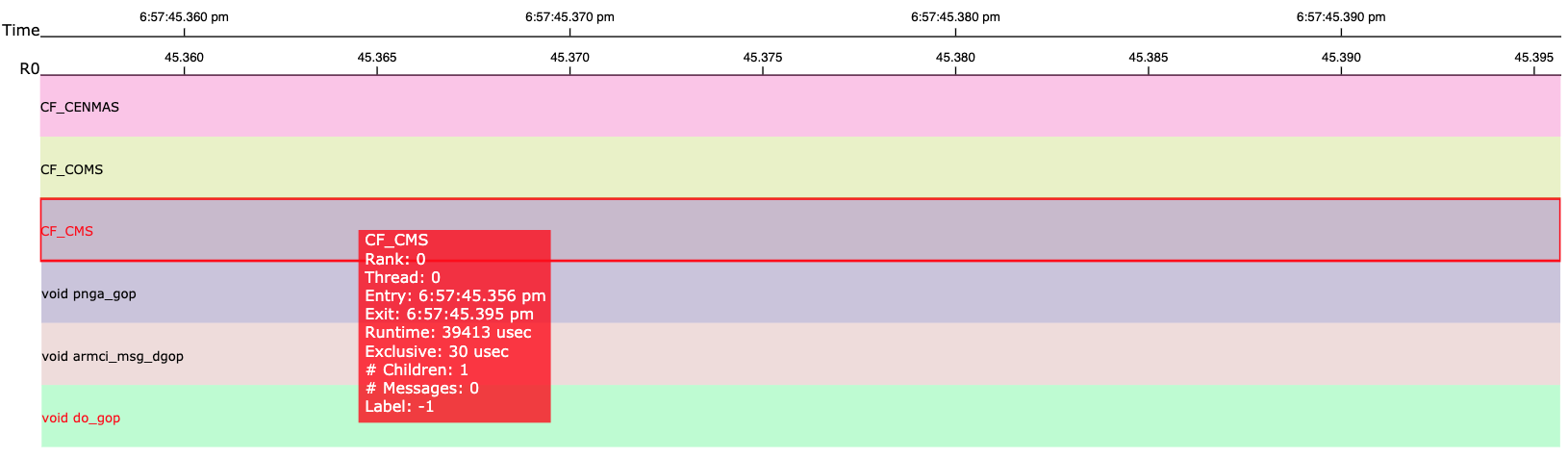}}
\caption{The case study showing anomalous function ``CF\_CMS."}
\label{fig:cf_cms}
\end{figure}

Then, the scientist switched to Rank 0 and wanted to check how ``MD\_FORCES" may affect other ranks. He found that Rank 0 mainly suffered anomalies in ``MD\_FINIT" (Fig.~\ref{fig:md_finit} and ``CF\_CMS" (Fig.~\ref{fig:cf_cms}). Both routines were related to the ``MD\_FORCES." Essentially, ``MD\_FINIT" initializes the calculation of the forces in every time step. Within that, ``CF\_CMS" calculates the center of mass of the solute atoms. This center of mass calculation involves calculating the center of mass of the solute atoms on every processor. A global sum is used to calculate the center of mass across all processors. Next, the distance squared of every solute atom to the center of mass is calculated and summed per processor. Another global sum calculates the sum of distance squared across all solute atoms. Taking the square root generates something of a standard deviation of the mass of the solute atoms relative to their center of mass. The scientist supposed Rank 0 had such problems here because it needed to be involved in the global sums, but it also had a unique role to participate in other functions which might cause it to fall behind other processors.

\begin{figure}[!b]
\centerline{\includegraphics[width=0.5\textwidth]{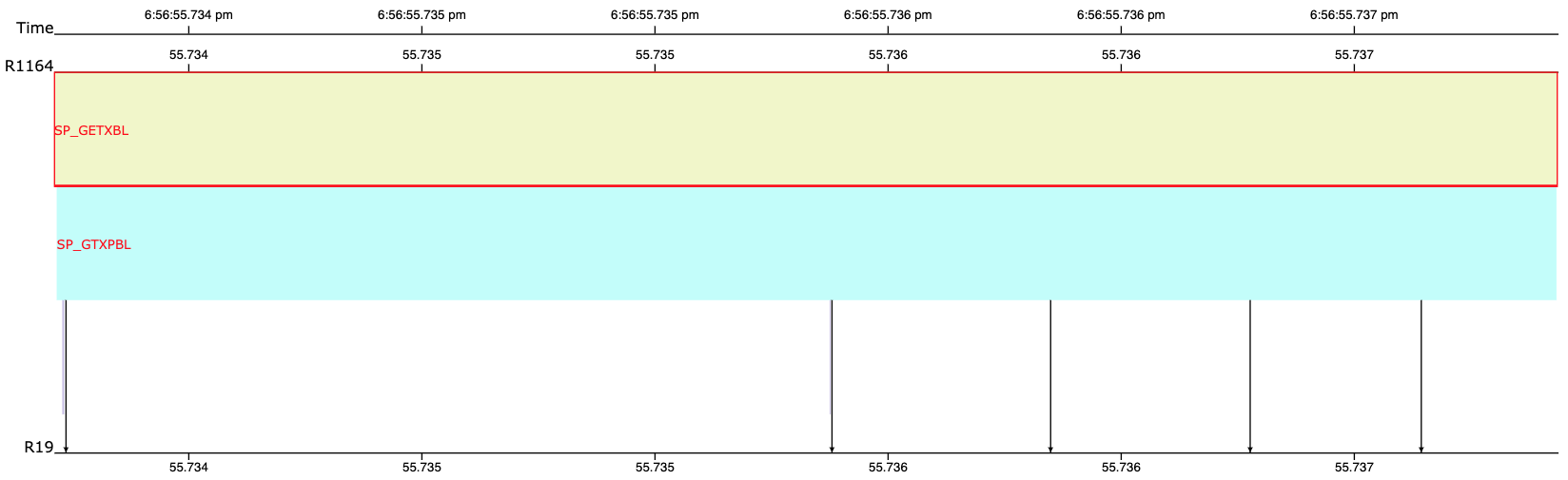}}
\caption{The case study showing anomalous function ``SP\_GETXBL."}
\label{fig:sp_getxbl}
\end{figure}

Another interesting finding drew the scientist's attention. He found that all other processes than Rank 0 tended to have anomalies in ``SP\_GTXPBL" (or the wrapper routine ``SP\_GETXBL," see Fig.~\ref{fig:sp_getxbl}). This routine fetches data about the atoms that live on other processes. The amount of data to fetch and subsequently process depends on the domain decomposition. When process $i$ fetches data from process $j$, it fetches the water (solvent) molecules separately from the solute atoms. However, depending on how the atoms are distributed across the domains, some of these get operations may take longer than others. Hence, this routine frequently gets flagged as taking an anomalously large amount of time. The scientist concluded that this observation helped him to understand the cause of problem by explaining it from the domain perspective.

\section{Literature Review}
Workflows are increasingly used for orchestrating the execution of complex scientific code suites in extreme-scale and highly heterogeneous environments. However, to date, we cannot reliably predict, understand, and optimize workflow performance. Sources of performance variability---particularly, the interdependencies of workflow design---execution environment, and system architecture are not well understood. While there is a rich portfolio of tools for performance analysis~\cite{Lammel2016,Shende1998,Knupfer2012,Tallent2008}, modeling, and prediction for single applications, these have not been adapted to handle workflows. Workflows have specific performance issues based on underlying resource management, potential contention for resources, and interdependence of the different workflow tasks. Performance tools for single applications, such as Score-p and TAU, produce similar output in the form of Event Trace and Profile files that can be read by performance and analysis tools, such as Jumpshot, Paraprof, Vampir, and others. The Profile file contains a concise summary of events, while the Event Trace includes event details with timestamps and can be quite large. Starting with Version 2.27, TAU has been modified to integrate with Chimbuko and now offers some support for workflows.  HPCToolkit~\cite{Tallent2008} does not provide the level of granularity needed to detect the interdependencies between resources inherent in workflows. Currently, there are no tools available apart from Chimbuko that capture performance provenance for single applications or workflows. Other tools, such as XALT~\cite{agrawal2014}, are used by facility administrators to track user environments and do not extract performance.

Numerous efforts aimed at reducing and better understanding detailed performance data exist, e.g., data compression, feature extraction, and performance modeling. However, these efforts are designed to operate post hoc on complete performance data sets rather than in real time and may not accommodate the level of details needed to address the complexity of emergent workflows. Furthermore, as these tools work post hoc, they require the collection of extreme-sized performance data files, severely limiting their applicability for complex applications and workflow scenarios.
Performance visualization is crucial for the representation and diagnosis of HPC application performance~\cite{isaacs2014state}~\cite{ivapp18}, showcasing different levels and aspects of the performance data. For trace events, visualizing the events along a time axis is an intuitive design as in Vampir~\cite{knupfer2008vampir} and Jumpshot~\cite{Zaki1999}. These temporal visualizations provide level-of-detail explorations, so users can zoom into different time window granularities~\cite{ivapp18}. Function invocation visualization is critical to understand the potential trigger for runtime problems. A directed tree or graph usually is employed to present the structure in a call stack, such as Vampir~\cite{knupfer2008vampir} and CSTree~\cite{Xie2018}. Message communication between functions also is important information to visualize, as well as a common reason for application latency. A directed line is drawn to represent the message delivery path, which is adopted in Jumpshot~\cite{Zaki1999}. Vampir also summarizes the communication between threads or processes in terms of an adjacency matrix~\cite{knupfer2008vampir}.

The major drawback of existing visualization methods is their limited capability for online performance evaluation and processing of streaming performance data in an \textit{in situ} fashion. However, online data reduction and sampling are critical for extreme-scale applications to cope with their tremendous performance data volumes.
Chimbuko alone provides scientific application developers with a single framework that integrates \textit{in situ} performance data reduction with provenance  while maintaining explainability thanks to its visualization module.

\section{Summary}
In this work, we presented Chimbuko, the performance analysis framework for real-time, distributed streaming AD and visualization. Our approach is a scalable performance framework that targets the capture and evaluation of trace-level performance data for scientific workflows at the exascale. It provides significant performance data reduction with a factor of 148 times---without adding unreasonable additional computational overhead. The streaming AD was shown to be comparable in its accuracy to other batch processing modules. Moreover, the visualization framework presented a scalable solution for backend data processing and frontend multiscale data representations. Finally, to showcase its effectiveness, we presented the results of a case study of the Chimbuko framework on Summit, successfully analyzing a NWChemEx workflow study. 

For future work, we will boost Chimbuko's overall performance 
to further minimize existing application overhead and maximize detection power of anomaly behaviors in trace data. As part of the effort to improve its AD capability, we will use a more advanced AD algorithm to extend the AD module. Furthermore, to achieve minimal overhead, we will seamlessly integrate the AD module and TAU. For visualization, we intend to adopt a more advanced distributed database for simultaneous writing and advanced visual design.
\section*{Acknowledgment}
This research was supported by the Exascale Computing Project (17-SC-20-SC), a collaborative effort of the U.S. Department of Energy Office of Science and the National Nuclear Security Administration.

%
\bibliography{chimbuko}
\bibliographystyle{plain}
%

\end{document}